\documentclass[10pt, conference]{IEEEtran}
\ifCLASSINFOpdf
\else
\fi
\usepackage{url}

\usepackage{tikz}
\usepackage{graphicx}
\usetikzlibrary{shapes,arrows,automata}

\usepackage{mathpartir}
\usepackage{amsmath,amsfonts,amsthm,amssymb,mathtools,thmtools}
\usepackage{mathrsfs}  
\usepackage{stmaryrd}
\usepackage{thm-restate}
\usepackage{environ}

\usepackage{listings}
\usepackage[inline]{enumitem}

\usepackage{prftree}
\usepackage{upgreek}
\usepackage{wasysym}
\usepackage{subcaption}
\usepackage{hyperref}

\usepackage{cancel}

\usepackage{xcolor}

\renewcommand{\tablename}{Table}
\renewcommand{\figurename}{Figure}

\lstdefinelanguage{MuAC}{
	keywords = {Gives, Me, with}
}

\lstdefinestyle{MuAC}{basicstyle=\footnotesize\ttfamily, keywordstyle=\color{gray}, language=MuAC}

\newtheorem{definition}{Definition}

\def\makenewenum#1#2{%
\newcounter{cnt#1}
\newenvironment{#1}%
{\begin{list}{\makebox[0pt][r]{#2}}%
{\setlength{\itemsep}{0pt}%
 \setlength{\parsep}{.2em}%
 \setlength{\leftmargin}{2em}%
 \setlength{\labelwidth}{.2em}%
 \usecounter{cnt#1}}}%
{\end{list}}}
\makenewenum{enu}{\arabic{cntenu}.}
\makenewenum{enum}{(\arabic{cntenum})}
\makenewenum{itenum}{\em(\arabic{cntenum})}
\makenewenum{en}{(\roman{cnten})}
\makenewenum{renum}{\textit{(\roman{cntrenum})}}
\makenewenum{enhyphen}{(\roman{cntenhyphen}')}
\makenewenum{aenum}{(\alph{cntaenum})}
\makenewenum{itemizes}{$\bullet$}

\newenvironment{restate-theorem}[1]%
  {\begin{trivlist}\item[]{\normalsize\bfseries{\sffamily}
        Restatement of Theorem~#1.}\hspace*{0mm}\it}%
  {\end{trivlist}}

\newenvironment{restate-lemma}[1]%
  {\begin{trivlist}\item[]{\normalsize\bfseries{\sffamily}
        Restatement of Lemma~#1.}\hspace*{0mm}\it}%
  {\end{trivlist}}

\newenvironment{restate-proposition}[1]%
  {\begin{trivlist}\item[]{\normalsize\bfseries{\sffamily}
        Restatement of Proposition~#1.}\hspace*{0mm}\it}%
  {\end{trivlist}}

\newcommand{\semdeno}[1]{\llbracket #1 \rrbracket}
\newcommand{\semden}[1]{\llparenthesis\, #1 \,\rrparenthesis}

\def\makenewenum#1#2{%
	\newcounter{cnt#1}
	\newenvironment{#1}%
	{\begin{list}{\makebox[0pt][r]{#2}}%
			{\setlength{\itemsep}{0pt}%
				\setlength{\parsep}{.2em}%
				\setlength{\leftmargin}{2em}%
				\setlength{\labelwidth}{.2em}%
				\usecounter{cnt#1}}}%
		{\end{list}}}

\NewEnviron{smalign*}{%
			\begin{align*}
			\BODY
			\end{align*}}

\newcommand{\linearcontract}{\multimap\hspace{-8pt}\multimap}

\newcommand{\MuACstate}{\mathfrak{S}}

\newcommand{\usr}{\mathfrak{u}}
\newcommand{\Usr}{\mathfrak{U}}
\newcommand{\Res}{\mathfrak{R}}
\newcommand{\res}{\mathfrak{r}}

\newcommand{\pol}{\mathfrak{P}}

\newcommand{\Pred}{\mathbb{P}}

\newcommand{\GiveLs}{\mathit{GiveLs}}

\begin{document}
	%
	\title{Automatic Fair Exchanges}

	
	
	\author{
	\IEEEauthorblockN{
	  Lorenzo Ceragioli\IEEEauthorrefmark{1},
	  Letterio Galletta\IEEEauthorrefmark{2},
	  Pierpaolo Degano\IEEEauthorrefmark{1},
	  and Luca Viganò\IEEEauthorrefmark{3}
	}
	\IEEEauthorblockA{\IEEEauthorrefmark{1} Universit\`a di Pisa, Italy}
	\IEEEauthorblockA{\IEEEauthorrefmark{2} IMT School for Advanced Studies Lucca, Italy}
	\IEEEauthorblockA{\IEEEauthorrefmark{3} Department of Informatics, King's College London, UK}
	}
	
%
%

	\maketitle

\begin{abstract}
In a decentralized environment, exchanging resources requires 
users to bargain
until an agreement is found.
Moreover, human agreements involve a combination of collaborative and selfish behavior and often induce circularity, complicating the evaluation of exchange requests.
We introduce MuAC, a policy language that allows users to state in isolation under which conditions they are open to grant their resources
and what they require in return.
In MuAC, exchange requests are evaluated automatically with the guarantee that the only exchanges that will take place are those that mutually satisfy users' conditions.
%
%
Moreover, MuAC can be used as an enforcement mechanism to prevent users from cheating.
As a proof of concept, we implement a blockchain smart contract that allows users to exchange their non-fungible tokens.
\end{abstract}


\section{Introduction}\label{sec:intro}

Exchanges of resources normally occur in many contexts, and human decisions are heavily based on reciprocity, as it is common for human interactions~\cite{HumanNature}.
As an example, consider the \emph{home exchange}, a common form of lodging in which two parties agree to swap homestays for a set period of time.
Say Alice 
offers to exchange her house in the countryside with Bob who owns a house downtown.
The decision is made by the two users based both on the properties of their houses and on their preferences.
%
However, more involved transactions may occur.
Consider, e.g., Bob's friend Carl, who owns a flat downtown and would like to spend a week in the countryside at Alice's house.
However, Alice does not like staying in a flat, so there is no direct agreement with Carl, but Bob can generously accept to ``pay for Carl'', giving Alice his house in place of Carl's flat.

This example illustrates how, when some user requests the resource of another, the two users, and possibly more, can interact to bargain until an agreement is found.
This interaction requires users to be online at the same time and to spend time bargaining, which is hard in a distributed environment.
One way to tackle this problem is to let users state in isolation the policies that specify the resources they are willing to offer \emph{together with} those they require in return.

%
For example, the policies of Alice, Bob and Carl may be as follows:
\begin{description}
	\item[Alice:] I will give my countryside house to get one downtown
	\item[Bob:] I will give my house downtown to get one in the countryside for me or a friend of mine
	\item[Carl:] I will give my flat downtown to get a house in the countryside
\end{description}

Each user states in isolation her own policy that only controls the access to her resources.
Clearly, the policies in force in a community impact on each of the participants. 
Thus, their composition has to be considered to verify whether an exchange request violates no policy and can thus be served.
Evaluating requests may induce some circularity.
For instance, Alice agrees to grant her house if Bob does the same, and vice versa.
Nevertheless, the agreement trivially satisfies the policies of both Alice and Bob.
Circularity may also have involved forms when more users come into play.

Here, we introduce MuAC, a policy language that allows users to state under which conditions they are open to grant their resources to others and what they require in return.
MuAC is equipped with a formal semantics in terms of a transition system. 
Transitions change the ownership of the resources in the system, but one has to check that a resource exchange obeys all the policies of the users involved, and for that a centralized authority is on demand.
Rather than directly relying on such a notion of exchange agreement, we compile MuAC policies to equivalent logical theories and compute agreements as proofs.

Since classical logic is not adequate to express the circular agreements that are typical of human contracts, we start from contractual logic~\cite{BZ} and build on~\cite{LNL}.
For that, we introduce a non-standard mixed logic composed of a linear fragment
(for reasoning about resources) and a non-linear one (for classical conditions, like relationships and attributes).

MuAC and its logic offer the basis of an enforcement mechanism that prevents users from cheating.
An implementation as a blockchain smart contract lets users define their policies and to exchange resources like non-fungible tokens (NFTs) with granted mutual advantage.
The usage of an off-chain client reduces the on-chain cost to be linear.

\paragraph*{Related work}
Mutuality plays a role in access control for cooperative systems, in particular in discretionary access control~\cite{CompSecPrinPra}.
In this case, a main issue is the combination of individual policies.
To the best of our knowledge, no proposals address mutuality, but only focus on the resolution of conflicts~\cite{policy-composition, collac, surveyCCCS}.
A remarkable exception is~\cite{SACMAT19} that permits defining mutual access control policies.
A new grant is introduced, called  \textit{mutual}, besides the usual \textit{accept} and \textit{deny}.
A first difference with our proposal is that in~\cite{SACMAT19} mutuality is hard-coded in the access control mechanisms, and is not user-defined.
In addition, MuAC considers also involved forms of agreements with many users, and targets finite resources.
In our previous work~\cite{ITAsec20} we proposed a policy language to state conditions about what a user receives in return for allowing an access.
Differently from this paper, in~\cite{ITAsec20} we did not consider finite resource exchanges but only data sharing, like, e.g., on social networks, and we proposed neither an implementation nor a formal semantics.

Mutuality plays a main role also in trust negotiations that
permit a safe interaction between two parties that do not trust each other~\cite{KolarGL18}.
Here too, each party defines an individual policy specifying the conditions that the other party must satisfy 
to obtain credentials.
Logical languages for specifying trust policies have been proposed, e.g.,
Cassandra~\cite{BeckerS04} and SecPal4P~\cite{secpal4p}, but these are based on classical logic, and thus circular conditions do not lead to an agreement.

\section{Defining Exchange Policies}\label{sec:col-MuAC}

First, we	present the notion of exchange environment to model systems where users want to share 
resources; then, we introduce the syntax of MuAC and we sketch its semantics.

%

\paragraph*{Exchange Environment}

We assume as fixed 
a set $\Res$ of \emph{resources} (ranged over by $\res, \res', \res''$)
and
a set $\Usr$ of \emph{users} (ranged over by $\usr, \usr', \usr''$).
An \emph{exchange environment} is a transition system with states $\MuACstate$ associating users with the (multiset of) resources they own, and transitions $\MuACstate \rightarrow \MuACstate'$ representing exchanges that change the ownership of resources.
A \emph{computation} from the state $\MuACstate$ to the state $\MuACstate'$ is the reflexive, transitive closure of $\rightarrow$, 
denoted by $\MuACstate \rightarrow^* \MuACstate'$.

\paragraph*{MuAC policy language}

Let $U$ be the set of user variables, ranged over by $u, u', u'', u_i$, with the distinguished element \texttt{Me},
and let $\Pred$ be the set of predicate symbols, ranged over by $p, p', p''$.
\begin{definition}[MuAC policies]\label{def:MuPol}
The MuAC policy $\pol_\usr$ of the user $\usr$ is a set of rules $\nu$ given by the following grammar:
\begin{align*}
\nu &::= \texttt{Gives}(\texttt{Me}, \res, u) \texttt{ :- } GiveLs \ \mathit{with}\  P \\ 
P &::= p(u_1, \dots , u_n)\ P \mid \epsilon \\
\GiveLs &::= \texttt{Gives}(u, \res, u')\ \GiveLs \mid \epsilon
\end{align*}
\end{definition}


Intuitively, \texttt{Me} refers to the policy owner, whereas all other user variables are universally quantified.
A rule states that the policy owner is willing to give a resource $\res$ to the requester $u$ if a list of properties $P$ holds (e.g., friendship), 
provided that the exchanges specified in \texttt{GivesLs} also take place.

For example, the MuAC policy $\pol_{Alice}$ of Alice is
\begin{lstlisting}[style=MuAC]
Gives(Me, countryside_house, u) :- 
    Gives(u', downtown_house, Me)
\end{lstlisting}
the MuAC policy $\pol_{Bob}$ of Bob is 
\begin{lstlisting}[style=MuAC]
Gives(Me, downtown_house, u) :- 
    Gives(u', countryside_house, u'') 
      with FriendOrSame(Me, u'')
\end{lstlisting}
and the MuAC policy $\pol_{Carl}$ of Carl is
\begin{lstlisting}[style=MuAC]
Gives(Me, downtown_flat, u) :- 
    Gives(u', countryside_house, Me)
\end{lstlisting}

The semantics $\semdeno{\pol_\usr}$ of a MuAC policy $\pol_\usr$ is the set of computations that are accepted by the user $\usr$, where we assume a \emph{context} $\Gamma$ to interpret the predicate symbols (e.g., to store that Bob and Carl are friends).
We filter the exchange environments computations taking only the fair ones, i.e., the ones that are accepted by all the users.

For example, assuming $\MuACstate_0$ to be the initial state where users are associated with their own houses, and $\MuACstate_{\usr, \usr'}$ to be a state in which a house exchange is performed between $\usr$ and $\usr'$, one can see that $\MuACstate_0 \rightarrow^* \MuACstate_{Alice, Bob}$ is fair, whereas $\MuACstate_0 \rightarrow^* \MuACstate_{Alice, Carl}$ is not accepted by Alice's policy.

\section{Enforcing MuAC}\label{sec:col-impl}
\paragraph*{Computing Fair Computations}\label{sec:col-formal}

We define the decidable non-standard logic MuACL, and prove that it precisely characterizes MuAC's semantics.
A \emph{MuACL theory} is a multiset of both linear and non-linear formulas.
Moreover, the new linear operator $\linearcontract$ is added, called \emph{linear contractual implication}, inspired by PCL~\cite{BZ}.
Roughly, a formula $\phi \linearcontract \phi'$ states that $\phi'$ will eventually hold provided that $\phi$ is \emph{true}.
Differently from common implication, $\phi \linearcontract \phi', \phi' \linearcontract \phi \vdash \phi \otimes \phi'$ ($\otimes$ is the standard multiplicative linear conjunction).
This new operator resolves circularity that often occurs in human reasoning.

We compile policies $\pol_\usr$ into MuACL theories.
Roughly, each rule $\texttt{Gives}(\texttt{Me}, \res, u) \texttt{ :- } GiveLs \ \mathit{with}\  P$ is translated as $\psi \rightarrow G(\phi \linearcontract \phi')$ where $\psi$ depends on $P$, $\phi$ on $GiveLs$, and $\phi'$ on $\texttt{Gives}(\texttt{Me}, \res, u)$.
An encoding is also given for the context $\Gamma$ and for states $\MuACstate$.
The compilation $\semden{\_}$ is correct, i.e., it characterizes MuAC fair computations:

\begin{restatable}{theorem}{Logiccorrectness}
	Given $\{\pol_\usr\}_{\usr \in \Usr}$, $\Gamma$, $\MuACstate$ and $\MuACstate'$,
	\begin{align*}
	\forall \usr \in \Usr .\ \MuACstate \rightarrow^* \MuACstate' \in \semdeno{\pol_\usr}\ \ \text{iff}\ \
	\bigcup_{\usr \in \Usr} \semden{\pol_\usr}, \semden{\Gamma}, \semden{\MuACstate} \vdash \semden{\MuACstate'}
	\end{align*}
\end{restatable}

\paragraph*{MuAC Smart Contract}

We implement the MuAC-based smart contract MuACSC for exchanging NFTs, associated with users in the contract's internal state.
MuACSC serves as a wallet: users can upload or withdraw NFTs, and can define their MuAC policies.
MuACSC supplies users a requested NFT, managing the needed fair exchanges.
Assume, e.g., that Alice wants an NFT $B$.
An off-chain client gathers users policies and computes the MuACL proof certifying that a fair computation exists from the current state to one where $B$ belongs to Alice.
The client sends the proof, if any, to the smart contract requesting the needed exchanges.
MuACSC checks the proof validity and updates the current state accordingly.
Note that the client performs the most expensive part of the computation, while only a small verification is  on-chain.
Reducing the computation cost of the contract is critical, because in most blockchains every executed instruction is payed by the requester using an in-block currency.

\paragraph*{Future Work}\label{sec:col-conclude}

We plan to extend MuAC to express policy updates like ``I will start trading $r$ for $r'$ if you start trading $r''$ for $r'''$.''
Another extension is to include negation, allowing the definition of conflicts of interest and embargo policies like ``I will give you $r$ if you give nothing to Carl.''

\bibliographystyle{IEEEtran}
\bibliography{references}

\end{document}